\documentclass[english,aps,prl,longbibliography,twocolumn]{revtex4-1}
\usepackage[T1]{fontenc}
\usepackage[latin9]{inputenc}
\usepackage{amsmath}
\usepackage{amssymb}
\usepackage{graphicx}
\usepackage{esint}

\makeatletter
\@ifundefined{textcolor}{}
{%
 \definecolor{BLACK}{gray}{0}
 \definecolor{WHITE}{gray}{1}
 \definecolor{RED}{rgb}{1,0,0}
 \definecolor{GREEN}{rgb}{0,1,0}
 \definecolor{BLUE}{rgb}{0,0,1}
 \definecolor{CYAN}{cmyk}{1,0,0,0}
 \definecolor{MAGENTA}{cmyk}{0,1,0,0}
 \definecolor{YELLOW}{cmyk}{0,0,1,0}
}

\usepackage{relsize}



\usepackage{babel}

\makeatother

\usepackage{babel}
\begin{document}

\title{Simple model for the Darwinian transition in early evolution}

\author{Hinrich Arnoldt,$^{a}$ Steven H. Strogatz$^{b}$ and Marc Timme$^{\ast,a,c}$}

\affiliation{$^{a}$Network Dynamics, Max Planck Institute for Dynamics and Self-Organization,
37077 Göttingen, Germany}

\affiliation{$^{b}$Department of Mathematics, Cornell University, Ithaca,
New York 14853, USA}

\affiliation{$^{c}$Institute for Nonlinear Dynamics, Faculty of Physics, Georg
August University Göttingen, 37077 Göttingen, Germany}

\affiliation{$^{\ast}$Corresponding author: timme@nld.ds.mpg.de}
\begin{abstract}
It has been hypothesized that in the era just before
the last universal common ancestor emerged, life on earth was fundamentally collective. 
Ancient life forms shared their genetic material freely through massive horizontal gene transfer (HGT). 
At a certain point, however, life made a transition to the modern era of individuality and vertical descent. 
Here we present a minimal model for this hypothesized ``Darwinian transition.'' 
The model suggests that HGT-dominated dynamics may have been
intermittently interrupted by selection-driven processes during which
genotypes became fitter and decreased their inclination toward HGT.
Stochastic switching in the population dynamics with three-point (hypernetwork)
interactions may have destabilized the HGT-dominated collective state and led to the emergence of vertical descent 
and the first well-defined species in early evolution.
A nonlinear analysis of a stochastic model dynamics covering key features
of evolutionary processes (such as selection, mutation, drift and HGT) supports
this view. Our findings thus suggest a viable route from early collective evolution 
to the start of individuality and vertical Darwinian evolution.\\
\\
\textbf{Keywords:} Evolutionary dynamics, population genetics, horizontal gene transfer, early evolution, emergence of the first species, hypernetwork dynamics, Darwinian threshold

\end{abstract}
\maketitle

\section{Introduction}

\subsection{The last universal common ancestor}

In the final chapter of ``On the Origin of Species'', Charles Darwin speculated that all life on earth may have descended from a  common ancestor. As he observed, ``all living things have much in common, in their chemical composition, their germinal vesicles, their cellular structure, and their laws of growth and reproduction. $\dots$ Therefore I should infer from analogy that probably all the organic beings which have ever lived on this earth have descended from some one primordial form, into which life was first breathed''~\cite{ClassicDarwin1859}.  

A century after Darwin, molecular biology provided new lines of circumstantial evidence for a universal common ancestor. All organisms were found to use the same molecule (DNA) for their genetic material, as well as a canonical look-up table (the genetic code) for translating nucleotide sequences into amino-acid sequences~\cite{nirenberg1963coding, Woese1967GeneticCode, knight2001rewiring}.  
Further clues came from cross-species comparisons of the molecules involved in the most fundamental processes of life, such as protein synthesis, core metabolism,  and the storage and handling of the genetic material. The first such analysis~\cite{BioWoese1977}, based on snippets of ribosomal RNA, provoked a revolution in our understanding of life's family tree~\cite{HGTWoese2000,ClassicCiccarelli2006,BioPace2012}.  It indicated that life is divided into three different domains: the Archaea, the Bacteria and the Eucarya~\cite{BioWoese1977,HGTWoese2000,BioPace2012}. Later studies using other molecular sequences placed
the root of the tree, corresponding to the last universal common ancestor, somewhere between the Bacteria and Archaea~\cite{BioGogarten1989,BioIwabe1989,BioTheobald2010, dagan2010genome, williams2013archaeal}, roughly $3.5-3.8$ billion years ago.  
The nature of the last universal common ancestor, however, remains unresolved: Was it prokaryotic or eukaryotic? Did it thrive in extreme or moderate temperatures? Was its genome based on RNA or DNA? For a review, see Ref.~\cite{glansdorff2008last} and references therein. 

\subsection{The era of collective evolution}

Our work in this paper was inspired by a conjecture proposed by Woese and his colleagues~\cite{WoeseFox_progenote1977, HGTWoese1998, HGTWoese2002, HGTWoese2005}. According to this conjecture, the last universal common ancestor was a community, not a single creature. It marked a turning point in the history of life: before it, evolution was collective and dominated by horizontal gene transfer; after it, evolution was Darwinian and dominated by vertical gene transfer. 

In Woese's scenario, life in the epoch leading up to the universal ancestor
was intensely communal. It was organized into loose-knit consortia of
protocells far simpler than the bacteria or archaea we know today. Woese and
Fox~~\cite{WoeseFox_progenote1977} called those hypothetical ancient life
forms ``progenotes.'' The term signifies that the coupling between genotype
and phenotype had not yet fully evolved, mainly because the process for
translating genes into proteins had not yet fully evolved either.  A
rudimentary form of translation existed, but it was ambiguous and hence had a
statistical character. Instead of producing a single protein, early translation produced a cloud of similar proteins. This ambiguity in protein synthesis in turn limited the specificity of \emph{all} the progenote's interactions. For example, lacking the large, complex proteins necessary for accurate copying and repair of the genetic material, the progenote's genome was tiny and subject to high mutation rates.

Progenotes  were not well-defined organisms as such, because they had no individuality and no long-term genetic pedigree. Their genes and component parts could come and go, being swapped in or out with other members of the community via horizontal transfer. But because biochemical innovations produced by any member of the community were available to all, evolution at this time was rapid---probably more rapid than at any time since. Selection acted on whole communities, not on individual progenotes. Those communities that were better at sharing their biochemical breakthroughs flourished. Out of this cauldron of evolutionary innovation, the universal genetic code and its translational machinery co-evolved, in response to the selective pressures favoring efficient sharing and interoperability. 

Vetsigian, Woese, and Goldenfeld~\cite{HGTVetsigian2006} confronted and constrained these speculations with mathematical analysis and computer simulations. Going beyond Woese's conjectures, they  probed early evolution scientifically by interpreting the available data on the genetic code. Their dynamical model for the evolution of the genetic code~\cite{HGTVetsigian2006} showed that a collective state of life is required to obtain the observed~\cite{haig1991quantitative, freeland1998genetic} statistical properties of the code, in particular, its simultaneous universality and optimality. A later study by Goldenfeld and colleagues~\cite{butler2009extreme} provided further evidence that only a collective state of life could have created the highly optimized code used by all life today
% the genetic code to its observed extent of ``one in a million''
~\cite{freeland1998genetic}.

\subsection{The Darwinian transition}

How did the era of collective evolution come to an end? Woese speculated that as the translation process began to improve, and as progenotic subsystems became increasingly complex and specialized, it would have become harder to find foreign parts compatible with them. Thus, horizontal gene transfer would have become increasingly difficult. The only possible modifications at this point would have come from within the progenote's lineage itself, through mutation and gene duplication. It was in this way that the progenotes would have made the Darwinian transition~\cite{HGTWoese2002, HGTWoese2005} to become ``genotes,'' i.e., life forms with a tight coupling between their nucleic acid genotypes and their protein phenotypes, and that could therefore evolve through the familiar Darwinian process of vertical descent.  

The model considered below is an attempt to explore, in mathematical terms,
how the Darwinian transition from the collective state to the modern era of
individuality might have taken place. Our approach shares with
Ref.~\cite{HGTVetsigian2006} the outlook that a dynamical systems calculation
should be devised to support or refute the hypotheses considered. Our results
lend support to the proposed collective state of
life~\cite{WoeseFox_progenote1977, HGTWoese1998, HGTWoese2002,
  HGTWoese2005,HGTVetsigian2006} by providing a potential mechanism for the exit from that state.

%Instead, phylogenetic sequencing of different
%groups of genes led to incongruities suggesting diverse possible shapes
%for phylogenetic branching~\cite{BioZillig1989,BioGogarten1989,BioKandler1994,HGTWoese1998,HGTDoolittle1999,BioDelsuc2005,BioPace2012}.
%The picture remains especially blurred at the root of the tree and
%it thus remains unclear which of the three branches of life arose
%first~\cite{HGTDoolittle1999,HGTDagan2006}. In particular, there
%are now different propositions for the possible shape of the tree,
%especially regarding its form close to its origin~\cite{HGTDagan2006}.

\subsection{Horizontal gene transfer}

Over the last decades more and more evidence has accumulated that, besides
selection, mutation, and drift~\cite{ClassicDrossel2001}, another process drives evolution: horizontal gene transfer. 
Here we briefly review the main points about horizontal gene transfer relevant to the mathematical model developed below.

While reproduction implies a vertical transfer of genes from one entity to the next
in the phylogenetic tree, there are also processes in which possibly unrelated
individuals exchange genetic material during their lifetimes, i.e., horizontally in the sense of the tree. 
This transfer of genes within one generation is consequently termed horizontal 
gene transfer (HGT)~\cite{HGTSyvanen1985,HGTSyvanen1994,BioDelsuc2005,HGTThomas2005} or lateral gene transfer.
It is now widely accepted that HGT is a fundamental driving force
of evolution~\cite{anderson1966possible, anderson1970evolutionary, sonea1988bacterial, HGTSyvanen1994,HGTKurland2003,HGTDagan2006, takeuchi2014horizontal}, and that its existence raises profound theoretical problems for evolutionary biology. 
For example, the longstanding problem of defining bacterial species
~\cite{sonea1988bacterial, cohan2002bacterial, lawrence2002gene, gevers2005re, konstantinidis2005genomic, staley2006bacterial} 
is due, in part, to the promiscuous use of HGT by bacteria.
A recent primer on horizontal gene transfer
and its potential for evolutionary processes in general is given in
\cite{HGTDoolittle2011}.

As discussed above, if HGT was rampant in the early stages of evolution, the last universal common ancestor was a community, not a single organism~\cite{HGTWoese1998,HGTWoese2002,HGTVetsigian2006,HGTGoldenfeld2007}. 
In this collective state, individuals could not yet be distinguished, as
each progenote's genes were frequently exchanged through HGT. In terms of the model to be developed below, the total pool of genotypes in the collective state 
would be spread out and thus broadly distributed in the state space of all theoretically possible genotypes. 
Conversely,  a genotype distribution that is highly localized in state space, being concentrated on just one or a few genotypes, 
would be the model's version of a well-defined species.

Woese postulated that as the
collective state of the progenote population slowly evolved toward higher complexity, its rate of HGT slowly decreased~\cite{HGTWoese2002}. 
At some point the system crossed the ``Darwinian threshold''~\cite{HGTWoese2002}. Then natural selection instead of HGT started to dominate the dynamics.
The fitter individuals were selected for and the first species emerged
from the distributed state. In the colorful language of Dyson \cite{HGTDysonNewYorkReview2007}: 
\begin{quote}
But then, one evil day, a cell resembling a primitive bacterium happened to find itself one jump ahead of its neighbors in efficiency. That cell, anticipating Bill Gates by three billion years, separated itself from the community and refused to share. Its offspring became the first species of bacteria---and the first species of any kind---reserving their intellectual property for their own private use. With their superior efficiency, the bacteria continued to prosper and to evolve separately, while the rest of the community continued its communal life. Some millions of years later, another cell separated itself from the community and became the ancestor of the archaea. Some time after that, a third cell separated itself and became the ancestor of the eukaryotes.
\end{quote}

After making a Darwinian transition,  evolution proceeds in the familiar vertical manner, being driven by selection, mutation, and drift~\cite{ClassicDrossel2001}, with
HGT playing only a minor role. Such Darwinian dynamics have, of course, been studied extensively in both experimental and  model settings~\cite{ClassicDrossel2001,ClassicTaylor2004,ClassicTraulsen2005,BasicsNowak2006,BioDurrett2008,ClassicArnoldt2012}. Compared to the dynamics of HGT their properties are relatively well understood.
Recently, potential influences of HGT on such evolutionary dynamics
have been investigated~\cite{HGTWoese1998,HGTVetsigian2006,HGTLeisner2008,MoranScience2010,MayerBMCEvolBiol2011,HGTVogan2011,HGTFuentes2014}.
Some mathematical models of HGT have focused on how it can increase
a population's fitness in Darwinian evolution~\cite{HGTWylie2010}. 

Keep in mind, however, that the hypothesized HGT associated with progenotes and the Darwinian transition, being associated with ribosomal genes 
and the rest of the core machinery of the cell, would have been of far greater evolutionary significance than the HGT of, say, antibiotic resistance 
genes seen in bacterial communities today. In Woese's scenario, the ancient form of HGT was rampant, pervasive, and tremendously disruptive and innovative. It was the prime mover in shaping the fabric of the cell~\cite{HGTWoese2005}. 

We would like to understand what such a Darwinian transition would look like, mathematically. 
The model described in the next section is deliberately minimal. It leaves out all the biology of ribosomes, proteins, genetic codes, and the like.
What remains is an attempt to capture the essence of Woese's speculations. In place of a community of progenotes, we 
consider a community of abstract genomes, represented by binary sequences. They interact via HGT, and are subject to mutation, selection, and drift on a fitness landscape. 
Our work suggests that HGT-dominated dynamics
may have been intermittently interrupted by selection-driven processes
during which genotypes became fitter and decreased their inclination
toward HGT. Such stochastic switching in the nonlinear population
dynamics may have destabilized the HGT-dominated state and thus 
led to a Darwinian transition and the emergence of the first species in early evolution. 

On a side note, an interesting mathematical aspect of the model is that it necessarily involves three-point interactions, 
since HGT transforms one genotype into a second by importing pieces of a third. 
Thus the model provides a natural biological example of a complex \emph{hypernetwork}. 
Until now, most models in evolutionary dynamics and population biology did not need to go beyond ordinary network structure, 
with two-point interactions between nodes connected by links.  

%%%%%%%%%%%%%%%%%%%%%%%%%%%%%%%%%%%%%%%%%%%%%%%%%%%%%%
\section{Evolutionary model}

To explore the consequences of HGT on evolution, we consider a model community of $N$ progenotes 
evolving on a fitness landscape~\cite{ClassicDrossel2001,ClassicWilke2001}
in the presence of selection, mutation, drift, and HGT. Each progenote carries
a genome of length $l$ composed of a sequence of the bases $0$ and
$1$. The genome of progenote $i$ determines its fitness $f_{i}$.
The progenotes reproduce by the Moran process~\cite{ModelMoran1962,ClassicDrossel2001},
i.e.,~each progenote reproduces randomly in time, with its reproduction
rate given by its fitness $f_{i}$. Whenever a progenote of genotype $i$ reproduces,
an offspring is added to the population which is either identical to genotype $i$ or a mutant of genotype $j$ with probability $\mu_{ij}$.
Instantaneously after such a reproduction event, one progenote in the population is
chosen randomly to die and is hence removed from the population.
We assume that one mutation event will only affect one of the bases of the genome, so that the Hamming distance between genotypes $i$ and $j$ is 1.

\begin{figure}[th]
\includegraphics[width=8cm]{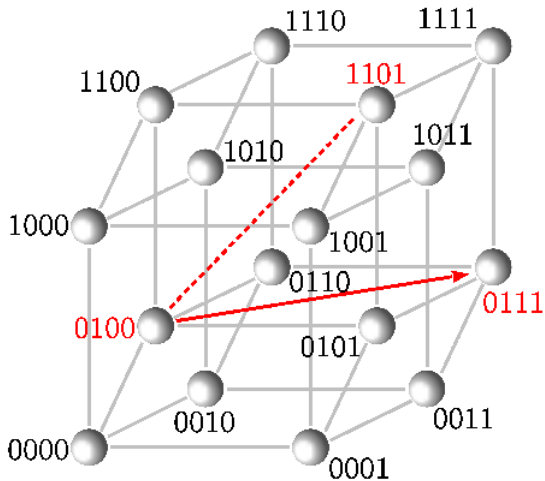}

\caption{\textbf{HGT-hyperlinks introduce three-genotype interactions to the
evolutionary dynamics, creating hypernetwork dynamics. }This schematic
example illustrates the insertion of an HGT-hyperlink (red solid and
dashed) to the sequence space of genomes of length $l=4$. Individuals
of genotype $A=0100$ take up the first two bases $\underline{11}$
from genotype $B=\underline{11}01$ which are inserted at position
three into $A=01\underline{11}00$. After the last two bits are cut
from the sequence, an progenote of genotype $A$ becomes of new genotype
$C=0111$.}

\label{fig:HGT-Link-Example}
\end{figure}

Hence, our fitness landscape may be represented by a network where
the different genotypes are the nodes of the network and the possible
mutations form the links. Assigning two different bases, 0 and
1, and given the structure of the mutations, the resulting network
is an $l$-dimensional hypercube 
(Fig.~\ref{fig:HGT-Link-Example}).

The fitness landscape underlying our model is assumed to be a Mount Fuji landscape~\cite{ClassicDrossel2001}: The highest fitness
is assigned to one single genotype, the peak. Other genotypes are assigned lower fitness:
the farther away from the peak in genotype space, the lower the fitness. Thus,
a single-peaked mountain landscape is created on genotype space, and a population evolving purely through the processes of selection and mutation should
converge to this peak. Note that the  Moran process described above is a
random process. It thereby constitutes a minimal model intrinsically
  including the effects of selection, mutation and genetic
  drift~\cite{ClassicDrossel2001}. The latter is induced by the stochastic
  selection in the combined process of reproduction, mutation and death and has the effect of randomly walking the population around in genotype space even if no fitness differences were present.

To reveal the potential impact of HGT we incorporate its basic features
into the stochastic evolution model. Two progenotes $A$ and $B$
may meet and a subsequence $s$ of progenote $B$'s genome may be
inserted into $A$'s genome. As a result of this horizontal gene transfer event, 
the genotype of progenote $A$ will transform into
another genotype $C$, determined by its original genotype and the
subsequence $s$. This process is illustrated in Figure \ref{fig:HGT-Link-Example}.

To model this process we add HGT-hyperlinks to the hypercube
network representing the fitness landscape. One such hyperlink symbolizes
a three-genotype interaction and is defined through the following
process. We choose two genotypes $A$ and $B$ randomly as well as
a random subsequence of genome $B$ with length between $x=2$ and
$x=l-2$ bases. This subsequence is inserted at a random position
of genome $A$. The remaining $x$ bases at the end of $A$'s sequence
are cut off, keeping the sequence length of $A$ constant. The new
sequence determines a genotype $C$, which genotype
$A$ becomes on interacting with $B$ via this HGT-link, denoted $(\overrightarrow{A,B,C})$.
If the resulting genotype $C$ is identical to $A$, this HGT-link
would not alter the population dynamics and would thus be irrelevant. 
We therefore neglect such self-projecting HGT-links.
We repeat the above procedure until a predefined number $m$ of new
HGT-links has been added to the system.

An HGT-link defines one type of HGT-event, in which part of
genotype $A$ is replaced by part of genotype $B$ and is thereby transformed to  
genotype $C$. We consider these events to occur independently
of each other. Let $k_X$ denote the number of progenotes of 
genotype $X$ in the population. Then the HGT-events above occur at a rate 
\begin{equation}
r_{A\rightarrow C}^{B}=c\cdot k_{A}\frac{k_{B}}{N}.\label{eq:HGTlink}
\end{equation}
Here the effective competence for HGT is modeled as a constant $c\geq0$ that
captures both the rate at which the progenotes meet and their actual preference
for the initiation of an HGT event, given that they meet. 

Note that interactions of the form~\eqref{eq:HGTlink},~independent of any
model details, imply collective dynamics on a complex hypernetwork,
due to their intrinsic three-genotype coupling involving $A$, $B$,
and $C$.  The dynamics of horizontal gene transfer in biological systems depends on a multitude
of factors, including the mode (e.g., natural transformation or conjugative
transfer) of HGT~\cite{HGTThomas2005}, and may vary with the fitness
of the donor and recipient~\cite{HGTLeisner2008,HGTLeisner2009} and other factors
such as environmental conditions \cite{HGTKovacs2009}. To focus on
qualitative mechanisms, we here consider the simplest setting where
$c$ is just a non-negative constant. We note that, via the factors
$k_{A}$, $k_{B}$ and the presence or absence of HGT-links
$(\overrightarrow{A,B,C})$, the actual rate of all HGT events in
the population still depends on how the population is distributed in genotype space.

\section{Quantifying Stochastic Switching}

To see how HGT influences the evolutionary dynamics we study how the
collective model dynamics depends on the competence $c$. The population sizes
$k_{i}(t)$ of progenotes of different genotypes~$i$ present in
the population fully describe the state of the system at time $t$.
We introduce the population entropy 
\begin{equation}
S(t)=-\sum_{i=0}^{2^{l}-1}\frac{k_{i}(t)}{N}\log\left[\frac{k_{i}(t)}{N}\right]\label{eq:PopEntr}
\end{equation}
 to quantify how broadly the population is distributed in genotype
space. Populations consisting of only one genotype have population
entropy zero. If the population is uniformly spread out in genotype
space, the population entropy takes its maximal value $S_{\text{max}}=l\log(2)$.

\begin{figure*}[t]
\includegraphics[width=15cm]{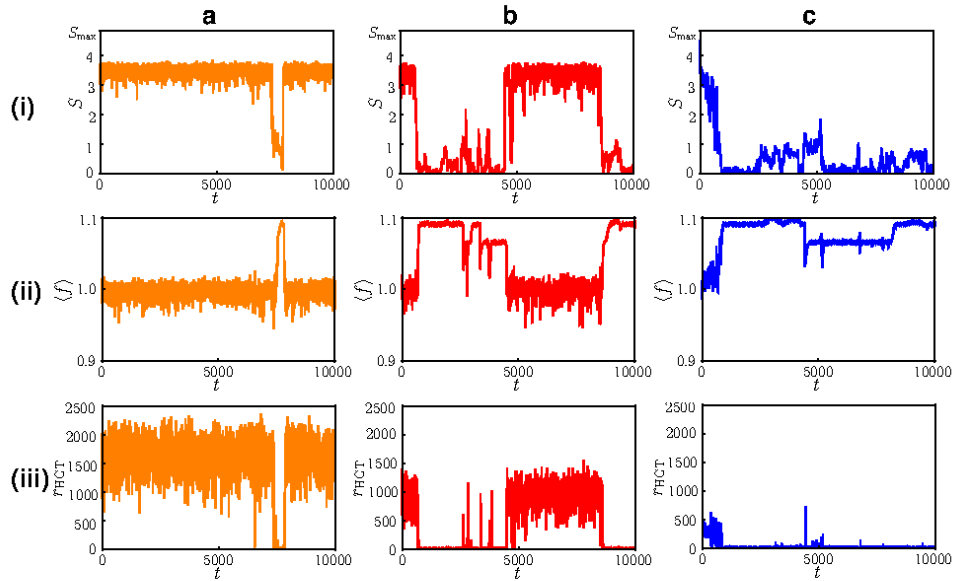} \caption{\textbf{The population dynamics is dominated by a speciated state
for low competence and a distributed state for high competence.} Shown
are example dynamics of the population entropy (\ref{eq:PopEntr})
for competence $c=5$ (a), $c=3$ (b) and $c=1$ (c) in an example
population of $N=1000$ progenotes with genome length $l=7$. (a): For
high competence $c$ the population entropy almost always fluctuates
around a high value for all initial conditions. (b): The dynamics switch
stochastically to a low entropy state and stay there longer for 
lower values of $c$.  (c): The low entropy state is rendered globally stable
for low $c$ so that the population entropy fluctuates slightly above
zero for all initial conditions. In the low entropy state the
population dynamics are driven by selection, in the high entropy state
by HGT. Panels (i) show the entropy dynamics, (ii) the average fitness
$\left\langle f\right\rangle $ of the population corresponding to
the entropy dynamics and (iii) the corresponding HGT rates $r_{\text{HGT}}$
that the population exhibits at time $t$. For low population entropies
the fitness is high and HGT rate small and vice versa for high population
entropies. The mutation probability was set to $\mu_{ij}=0.0001$.
Into the resulting Fujiyama fitness landscape~\cite{ClassicDrossel2001}
with fitness values between $f_{\text{min}}=0.9$ and $f_{\text{max}}=1.1$
we inserted $m=2000$ HGT-links.}
\label{fig:ExampleDynamics} 
\end{figure*}

Direct simulations of the stochastic dynamics reveal that for large
competences $c$, the collective dynamics converge to a state of high
population entropy where the population is highly spread out in genotype
space (Figure~\ref{fig:ExampleDynamics}a). It may only transiently
switch to a state localized in state space, i.e.,~with relatively
little spread in genetic material. In this high entropy state the total
HGT rate in the population is orders of magnitude higher than in a speciated
state (see below). The population does not adapt to the underlying fitness
landscape; in that sense, HGT is the main driving process in this
large-$c$ scenario. We identify this state of high population entropy
with a pre-Darwinian collective state, as in this state no distinct
species can be distinguished and HGT is the dominant force driving
the evolutionary dynamics.

In contrast, if the progenotes' competence for HGT is low, we observe
a population dynamics which converges toward a state of low population
entropy (Figure~\ref{fig:ExampleDynamics}c). This confirms the
observation that selection, mutation and drift will drive a population to
adapt to a fitness landscape if the mutation rate is not too high~\cite{ClassicDrossel2001,BasicsNowak2006}.
The population is thus concentrated around the fittest genotype
with only rare mutations and genetic drift causing some spread of the
population. As a consequence, large parts of the population exhibit
  the same or similar genotypes such that it is in a speciated state.

While the system spends almost all time close to its speciated state for
low competence, the dynamics switch stochastically between the speciated
and the distributed state if the competence is not small enough. 
Figures~\ref{fig:ExampleDynamics}a-c show that the higher the progenotes' competence for HGT is, 
the longer the system stays in the distributed state.

We conclude that both the speciated state and the distributed state
are dynamically accessible metastable states (for high enough competence).
The dynamics only switch from one of these states to the other due to rare
events in the stochastic dynamics. This is supported by the fact that
the dynamics switch between these states on much shorter time scales
than the time they remain in them (see also Figure
\ref{fig:HGTSmallMut} below).

\begin{figure}[h]
\includegraphics[width=6cm]{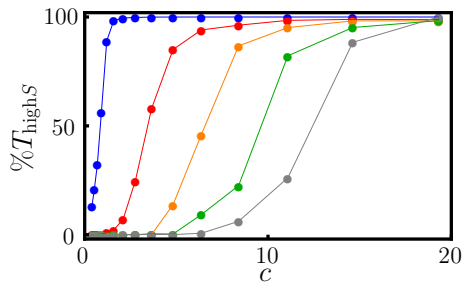} \caption{\textbf{The high entropy state is dynamically stable also for vanishing
mutation probabilities}. Shown is the measured percentage of time
a population stayed in the distributed state for a system with mutation
probabilities $\mu_{ij}=10^{-3}$ (blue), $\mu_{ij}=10^{-4}$, $\mu_{ij}=10^{-5}$
(orange), $\mu_{ij}=10^{-6}$ (green) and $\mu_{ij}=0$ (gray). Qualitatively,
the results are similar, only for higher mutation probabilities the
critical transition occurs at a lower value $c_{\text{cr}}$. System
parameters were $l=7$, $N=1000$ and $m=3000$ HGT-links were introduced
into a Fujiyama fitness landscape with fitness values between $f_{\text{min}}=0.9$
and $f_{\text{max}}=1.1$. Each datapoint was obtained in a simulation
of length $T=10^{6}$ with the initial condition $S(0)=S_{\max}$.}
\label{fig:HGTSmallMut}
\end{figure}

\begin{figure}[h]
\includegraphics[width=9cm]{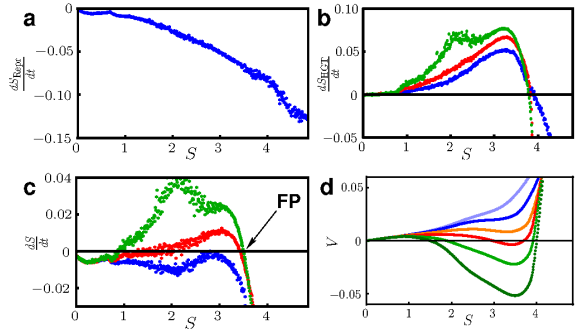} \caption{\textbf{The distributed state disappears for low HGT competence. }At
a critical competence the dynamical fixed point at high population
entropy is destroyed in a saddle-node bifurcation. Here we show the
analysis of the system yielding the dynamics illustrated in Figure~\ref{fig:ExampleDynamics}.
Panels (a) and (b) show the rate of change of the population entropy
due to reproduction and HGT obtained with the methods described in
the Methods section. The colors indicate the rate of change for competence
values $c=1$ (blue), $c=3$ (red) and $c=5$ (orange). Adding the
results from (a) and (b) according to equation (\ref{eq:SChange})
yields the overall rate of change $\dot{S}$ for the dynamics shown
in (c). The arrow indicates the fixed point at high population entropies
emerging through a saddle-node bifurcation for increasing the parameter
$c$. With equation (\ref{eq:HGTPotential}) we define a potential
$V(S)$ for the dynamics which is shown in (d) for the competences
$c=1$ (blue), $c=3$ (red) and $c=5$ (orange) and additionally for
$c=0.5$ (gray), $c=2$ (green) and $c=4$ (black). The potential
valley at high population entropies emerges between $c=1$ and $c=2$
so that the critical competence must lie between these two values.
Each dataset was obtained in simulations measuring the dynamics for
a time $T=10^{7}$.}
\label{fig:Forces} 
\end{figure}

How does the distributed state disappear for low competences? To answer
this question we developed a method based on the population entropy
defined in (\ref{eq:PopEntr}) to study the forces induced on the
dynamics by reproduction and HGT. The evolution dynamics are event-driven, 
and the population entropy $S$ may only change at these event
times. At each event there is a population entropy $S_{-}$ directly
before the event and a population entropy $S_{+}$ directly after
the event. The change of population entropy 
\begin{equation}
\Delta S=S_{+}-S_{-}
\end{equation}
induced by a single event will in general depend on the type of event
(reproduction or HGT) and the actual distribution of the population
over genotype space. If the population is in a state with population
entropy $S$, one event will thus induce a mean change $\overline{\Delta S}(S)$
averaged over all events occurring at population entropy $S$. The
rate $r(S)$ at which these events occur depends on the state of the
system as well. Multiplying the mean change induced by the single
events with the rate at which the events occur, we obtain the average
rate of change 
\begin{equation}
\frac{dS}{dt}=r(S)\cdot\overline{\Delta S}(S)
\end{equation}
induced on the dynamics. The reproduction and HGT events in our model
occur independently of each other, so that their contributions separate
additively according to

\begin{eqnarray}
\frac{dS}{dt} & = & \frac{dS_{\text{Repr}}}{dt}+\frac{dS_{\text{HGT}}}{dt}\\
 & = & r_{\text{Repr}}(S)\cdot\overline{\Delta S}_{\text{Repr}}(S)+r_{\text{HGT}}(S)\cdot\overline{\Delta S}_{\text{HGT}}(S).\label{eq:SChange}
\end{eqnarray}
We measured these functional dependencies in simulations of the dynamics
(for more details see Supplementary Information), thereby obtaining the forces
induced by reproduction and HGT which drive the population entropy
dynamics (Figure~\ref{fig:Forces}). The bistability of the dynamics
emerges because the impact of HGT increases with the diversity of
the population. Thus, if the population entropy is high, HGT will drive
it toward even higher population entropy, and hence toward the distributed state.
However, if the competence drops below a critical value, the impact
of HGT on the population's dynamics is always smaller than that of
selection, independent of the diversity of the population. Thus, the
distributed state disappears in a saddle-node bifurcation and the
population converges to a speciated state. Furthermore, our analysis
reveals that HGT alone can drive a population into a distributed state,
even in a total absence of mutations (Figure~\ref{fig:HGTSmallMut}).

Why do the dynamics almost always remain in the high entropy state
for high competence? Using the average rate of change~$\dot{S}(S)$
we define a potential 
\begin{equation}
V(S)=-\int_{0}^{S}\dot{S}(S')dS'\label{eq:HGTPotential}
\end{equation}
 in which the dynamics move under additional stochastic forcing. This
potential is shown in Figure~\ref{fig:Forces}d. According to reaction
rate theory~\cite{ChemistryHaenggi1990}, the depths of the two stable
states' potential wells determine the average time the dynamics stay
close to each of the stable states. As the potential well at the distributed
state becomes ever deeper for higher competence the dynamics hence
stay ever longer in this state.

Thus, our results suggest that when progenotes had high competence for
HGT in early evolution, a distributed state was dynamically stable.
If competence then decreased below a critical value, the distributed
state may have disappeared and triggered the emergence of the first species. For this
Darwinian transition to occur, the population's competence must have decreased
dynamically in the distributed state. How could this have happened?

\begin{figure}[h]
\includegraphics[width=6cm]{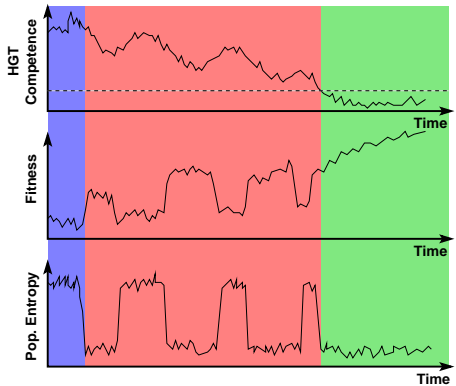} \caption{\textbf{A possible scenario for the evolution of distinct species
from a pre-Darwinian distributed state. }The three time series sketched
here are not simulation data, but encapsulate the speculations in the text, showing 
how the average competence, the average fitness, and the population
entropy may evolve in the transition from a distributed state to the
first distinct species. In the initial state (marked in blue) the
competence is high, so that HGT drives the dynamics; the population
exhibits a high population entropy and low average fitness. Through
a stochastic switching the dynamics reaches a state of low population
entropy where the fitness is higher as the population adapts to the
fitness landscape. Here the population could evolve slowly toward
lower competence. Thus, the dynamics switch back and forth between
the low and the high entropy state remaining longer and longer in
the low entropy state as the competence decreases (marked in red).
When the competence goes below a critical value (marked by the dashed
line in the top panel) the high entropy state disappears (marked in
green), the dynamics remains in the low entropy state, the population's
average fitness increases and the first species may robustly evolve.}
\label{fig:PrimordialSoupIdea}
\end{figure}

A decrease in competence may rely on a mechanism that combines the
stochastic switching uncovered above with the suggestion that fitter
populations may tend to be less prone to HGT events, 
as schematically illustrated in Figure~\ref{fig:PrimordialSoupIdea}. 
As the speciated state is always stable, even if the population's
competence is high, the population dynamics will stochastically switch
to this state repeatedly for relatively short times. In the selection-dominated
state (i.e., at low $S$) the population's fitness increases. A fitter
population that might be less prone to HGT events, as suggested recently
\cite{HGTVogan2011}, has a decreased overall competence (lower $c$ in
our simplified model setting). Smaller $c$ in turn increases the
stochastic residence times the population spends in the selection-dominated
state. This combination of two mutually amplifying contributions (decreasing
HGT rate and increasing fitness in the population) may yield decreasing
competence in the long term such that after sufficiently many switches
to the low-$S$ state, the competence may drop below a critical value
where the distributed state disappears. The population then stays
   localized around the fittest genotypes, thus marking the time of
  transition to Darwinian evolution. At this time, the first species
can robustly emerge. The scenario shown schematically in Figure~\ref{fig:PrimordialSoupIdea} illustrates
one potential course of such repeated switching dynamics, with temporarily
increased phases of higher fitness and decreasing HGT competence on
long time scales.

\section{Conclusion}

Our results provide a first glimpse of the possible dynamics that
may have led to the emergence of the first species from a distributed
state dominated by HGT. We demonstrated that a high competence for
HGT in a population may suffice to drive the population into a distributed
state (Figure~\ref{fig:ExampleDynamics}). In this state HGT
dominates the dynamics, in the sense that it inhibits the population's ability
to adapt to the underlying fitness landscape and thus prevents it from crystallizing into distinct
species. Our analysis revealed that, independently of the mutation
rate exhibited by the population, HGT can drive the population dynamics
into a state where the population is widely spread out in genotype
space (Figure~\ref{fig:HGTSmallMut}). We identify this state
with a pre-Darwinian collective state envisioned by Woese~\cite{WoeseFox_progenote1977, HGTWoese1998, HGTWoese2002, HGTWoese2005}.

Similarly, a state where no distinct species exist can emerge if the
mutation rate in the population is too large~\cite{BasicsNowak2006,ClassicEigen1971,ClassicEigen1979}.
Above a critical mutation rate (the error threshold) the population
cannot adapt to the underlying fitness landscape and will always evolve
toward a quasispecies
state~\cite{BasicsNowak2006,ClassicEigen1971,ClassicEigen1979} 
similar to the distributed state induced by HGT shown
above. However, there is a fundamental difference in the dynamics
induced by HGT and that induced by mutations: While a mutation rate
above an error threshold will always lead to a quasispecies state~\cite{BasicsNowak2006},
high rates of HGT as studied above induce a \emph{bistability} of
the dynamics where the distributed state coexists with a localized
``speciated'' state of low $S$. This coexistence may be essential
for the evolution toward lower competence in a population and thus
for the emergence of the first species; the coexistence is what enables a population
originally in a distributed state to repeatedly switch to a low-$S$
state. As selection plays a major role in such a low-$S$ state, progenotes
with lower competence would be selected for. Thus, with time, the
entire population would evolve toward lower competence until the
distributed state disappears as selection effects dominate the dynamics
and the first species emerge.

For the breakdown of the distributed state it is essential that the
population evolves toward a lower competence. That the latter may
in principle be possible was already suggested by Vogan and Higgs~\cite{HGTVogan2011}.
Our results on an idealized model now demonstrate how stochastic switching
and fitness-dependent competence may combine to create a transition
from a bistable state to a speciated-only state. They in particular
also suggest that HGT may be present at similar competence levels
before and after the emergence of the first species. From a complementary
perspective, whereas one or a few species may already have existed,
other population parts may still have been mixed without any clear
species. So the very first species may only have marked the beginning of
the decline of genuinely non-specific life, with other Darwinian transitions to follow.

\emph{Acknowlegdements:} We thank Nigel Goldenfeld, Stefan Grosskinsky,
Oskar Hallatschek and Arne Traulsen for valuable comments and discussions.
Supported by a grant of the Max Planck Society to MT. 

%\bibliographystyle{model2-names}
%\bibliography{References4}
\bibliography{References5}

%merlin.mbs apsrev4-1.bst 2010-07-25 4.21a (PWD, AO, DPC) hacked
%Control: key (0)
%Control: author (0) dotless jnrlst
%Control: editor formatted (1) identically to author
%Control: production of article title (0) allowed
%Control: page (1) range
%Control: year (0) verbatim
%Control: production of eprint (0) enabled
\begin{thebibliography}{59}%
\makeatletter
\providecommand \@ifxundefined [1]{%
 \@ifx{#1\undefined}
}%
\providecommand \@ifnum [1]{%
 \ifnum #1\expandafter \@firstoftwo
 \else \expandafter \@secondoftwo
 \fi
}%
\providecommand \@ifx [1]{%
 \ifx #1\expandafter \@firstoftwo
 \else \expandafter \@secondoftwo
 \fi
}%
\providecommand \natexlab [1]{#1}%
\providecommand \enquote  [1]{``#1''}%
\providecommand \bibnamefont  [1]{#1}%
\providecommand \bibfnamefont [1]{#1}%
\providecommand \citenamefont [1]{#1}%
\providecommand \href@noop [0]{\@secondoftwo}%
\providecommand \href [0]{\begingroup \@sanitize@url \@href}%
\providecommand \@href[1]{\@@startlink{#1}\@@href}%
\providecommand \@@href[1]{\endgroup#1\@@endlink}%
\providecommand \@sanitize@url [0]{\catcode `\\12\catcode `\$12\catcode
  `\&12\catcode `\#12\catcode `\^12\catcode `\_12\catcode `\%12\relax}%
\providecommand \@@startlink[1]{}%
\providecommand \@@endlink[0]{}%
\providecommand \url  [0]{\begingroup\@sanitize@url \@url }%
\providecommand \@url [1]{\endgroup\@href {#1}{\urlprefix }}%
\providecommand \urlprefix  [0]{URL }%
\providecommand \Eprint [0]{\href }%
\providecommand \doibase [0]{http://dx.doi.org/}%
\providecommand \selectlanguage [0]{\@gobble}%
\providecommand \bibinfo  [0]{\@secondoftwo}%
\providecommand \bibfield  [0]{\@secondoftwo}%
\providecommand \translation [1]{[#1]}%
\providecommand \BibitemOpen [0]{}%
\providecommand \bibitemStop [0]{}%
\providecommand \bibitemNoStop [0]{.\EOS\space}%
\providecommand \EOS [0]{\spacefactor3000\relax}%
\providecommand \BibitemShut  [1]{\csname bibitem#1\endcsname}%
\let\auto@bib@innerbib\@empty
%</preamble>
\bibitem [{\citenamefont {Darwin}(1859)}]{ClassicDarwin1859}%
  \BibitemOpen
  \bibfield  {author} {\bibinfo {author} {\bibfnamefont {C.~R.}\ \bibnamefont
  {Darwin}},\ }\href@noop {} {\emph {\bibinfo {title} {On the Origin of
  Species}}}\ (\bibinfo  {publisher} {John Murray},\ \bibinfo {address}
  {London, UK},\ \bibinfo {year} {1859})\BibitemShut {NoStop}%
\bibitem [{\citenamefont {Nirenberg}\ \emph {et~al.}(1963)\citenamefont
  {Nirenberg}, \citenamefont {Jones}, \citenamefont {Leder}, \citenamefont
  {Clark}, \citenamefont {Sly},\ and\ \citenamefont
  {Pestka}}]{nirenberg1963coding}%
  \BibitemOpen
  \bibfield  {author} {\bibinfo {author} {\bibfnamefont {MW}~\bibnamefont
  {Nirenberg}}, \bibinfo {author} {\bibfnamefont {OW}~\bibnamefont {Jones}},
  \bibinfo {author} {\bibfnamefont {P}~\bibnamefont {Leder}}, \bibinfo {author}
  {\bibfnamefont {BFC}\ \bibnamefont {Clark}}, \bibinfo {author} {\bibfnamefont
  {WS}~\bibnamefont {Sly}}, \ and\ \bibinfo {author} {\bibfnamefont
  {S}~\bibnamefont {Pestka}},\ }\bibfield  {title} {\enquote {\bibinfo {title}
  {On the coding of genetic information},}\ }in\ \href@noop {} {\emph {\bibinfo
  {booktitle} {Cold Spring Harbor Symposia on Quantitative Biology}}},\
  Vol.~\bibinfo {volume} {28}\ (\bibinfo {organization} {Cold Spring Harbor
  Laboratory Press},\ \bibinfo {year} {1963})\ pp.\ \bibinfo {pages}
  {549--557}\BibitemShut {NoStop}%
\bibitem [{\citenamefont {Woese}(1967)}]{Woese1967GeneticCode}%
  \BibitemOpen
  \bibfield  {author} {\bibinfo {author} {\bibfnamefont {C.~R.}\ \bibnamefont
  {Woese}},\ }\href@noop {} {\emph {\bibinfo {title} {The Genetic Code}}}\
  (\bibinfo  {publisher} {Harper and Row},\ \bibinfo {address} {New York},\
  \bibinfo {year} {1967})\BibitemShut {NoStop}%
\bibitem [{\citenamefont {Knight}\ \emph {et~al.}(2001)\citenamefont {Knight},
  \citenamefont {Freeland},\ and\ \citenamefont
  {Landweber}}]{knight2001rewiring}%
  \BibitemOpen
  \bibfield  {author} {\bibinfo {author} {\bibfnamefont {Robin~D}\ \bibnamefont
  {Knight}}, \bibinfo {author} {\bibfnamefont {Stephen~J}\ \bibnamefont
  {Freeland}}, \ and\ \bibinfo {author} {\bibfnamefont {Laura~F}\ \bibnamefont
  {Landweber}},\ }\bibfield  {title} {\enquote {\bibinfo {title} {Rewiring the
  keyboard: evolvability of the genetic code},}\ }\href@noop {} {\bibfield
  {journal} {\bibinfo  {journal} {Nature Reviews Genetics}\ }\textbf {\bibinfo
  {volume} {2}},\ \bibinfo {pages} {49--58} (\bibinfo {year}
  {2001})}\BibitemShut {NoStop}%
\bibitem [{\citenamefont {Woese}\ and\ \citenamefont
  {Fox}(1977{\natexlab{a}})}]{BioWoese1977}%
  \BibitemOpen
  \bibfield  {author} {\bibinfo {author} {\bibfnamefont {C.~R.}\ \bibnamefont
  {Woese}}\ and\ \bibinfo {author} {\bibfnamefont {G.~E.}\ \bibnamefont
  {Fox}},\ }\bibfield  {title} {\enquote {\bibinfo {title} {Phylogenetic
  structure of the prokaryotic domain: The primary kingdoms},}\ }\href@noop {}
  {\bibfield  {journal} {\bibinfo  {journal} {Proceedings of the National
  Academy of Sciences}\ }\textbf {\bibinfo {volume} {74}},\ \bibinfo {pages}
  {5088--5090} (\bibinfo {year} {1977}{\natexlab{a}})}\BibitemShut {NoStop}%
\bibitem [{\citenamefont {Woese}(2000)}]{HGTWoese2000}%
  \BibitemOpen
  \bibfield  {author} {\bibinfo {author} {\bibfnamefont {C.~R.}\ \bibnamefont
  {Woese}},\ }\bibfield  {title} {\enquote {\bibinfo {title} {Interpreting the
  universal phylogenetic tree},}\ }\href@noop {} {\bibfield  {journal}
  {\bibinfo  {journal} {Proc. Natl. Acad. Sci. USA}\ }\textbf {\bibinfo
  {volume} {97}},\ \bibinfo {pages} {8392--8396} (\bibinfo {year}
  {2000})}\BibitemShut {NoStop}%
\bibitem [{\citenamefont {Ciccarelli}\ \emph {et~al.}(2006)\citenamefont
  {Ciccarelli}, \citenamefont {Doerks}, \citenamefont {von Mering},
  \citenamefont {Creevey}, \citenamefont {Snel},\ and\ \citenamefont
  {Bork}}]{ClassicCiccarelli2006}%
  \BibitemOpen
  \bibfield  {author} {\bibinfo {author} {\bibfnamefont {F.~D.}\ \bibnamefont
  {Ciccarelli}}, \bibinfo {author} {\bibfnamefont {T.}~\bibnamefont {Doerks}},
  \bibinfo {author} {\bibfnamefont {C.}~\bibnamefont {von Mering}}, \bibinfo
  {author} {\bibfnamefont {C.~J.}\ \bibnamefont {Creevey}}, \bibinfo {author}
  {\bibfnamefont {B.}~\bibnamefont {Snel}}, \ and\ \bibinfo {author}
  {\bibfnamefont {P.}~\bibnamefont {Bork}},\ }\bibfield  {title} {\enquote
  {\bibinfo {title} {Toward automatic reconstruction of a highly resolved tree
  of life},}\ }\href@noop {} {\bibfield  {journal} {\bibinfo  {journal}
  {Science}\ }\textbf {\bibinfo {volume} {311}},\ \bibinfo {pages} {1283--1287}
  (\bibinfo {year} {2006})}\BibitemShut {NoStop}%
\bibitem [{\citenamefont {Pace}\ \emph {et~al.}(2012)\citenamefont {Pace},
  \citenamefont {Sapp},\ and\ \citenamefont {Goldenfeld}}]{BioPace2012}%
  \BibitemOpen
  \bibfield  {author} {\bibinfo {author} {\bibfnamefont {N.~R.}\ \bibnamefont
  {Pace}}, \bibinfo {author} {\bibfnamefont {J.}~\bibnamefont {Sapp}}, \ and\
  \bibinfo {author} {\bibfnamefont {N.}~\bibnamefont {Goldenfeld}},\ }\bibfield
   {title} {\enquote {\bibinfo {title} {Phylogeny and beyond: {Scientific},
  historical, and conceptual significance of the first tree of life},}\
  }\href@noop {} {\bibfield  {journal} {\bibinfo  {journal} {Proc. Natl. Acad.
  Sci. USA}\ }\textbf {\bibinfo {volume} {109}},\ \bibinfo {pages} {1011--1018}
  (\bibinfo {year} {2012})}\BibitemShut {NoStop}%
\bibitem [{\citenamefont {Gogarten}\ \emph {et~al.}(1989)\citenamefont
  {Gogarten}, \citenamefont {Kibak}, \citenamefont {Dittrich}, \citenamefont
  {Taiz}, \citenamefont {Bowman}, \citenamefont {Bowman}, \citenamefont
  {Manolsen}, \citenamefont {Poole}, \citenamefont {Date}, \citenamefont
  {Oshima}, \citenamefont {Konishi}, \citenamefont {Denda},\ and\ \citenamefont
  {Yoshida}}]{BioGogarten1989}%
  \BibitemOpen
  \bibfield  {author} {\bibinfo {author} {\bibfnamefont {J.~P.}\ \bibnamefont
  {Gogarten}}, \bibinfo {author} {\bibfnamefont {H.}~\bibnamefont {Kibak}},
  \bibinfo {author} {\bibfnamefont {P.}~\bibnamefont {Dittrich}}, \bibinfo
  {author} {\bibfnamefont {L.}~\bibnamefont {Taiz}}, \bibinfo {author}
  {\bibfnamefont {E.~J.}\ \bibnamefont {Bowman}}, \bibinfo {author}
  {\bibfnamefont {B.~J.}\ \bibnamefont {Bowman}}, \bibinfo {author}
  {\bibfnamefont {M.~F.}\ \bibnamefont {Manolsen}}, \bibinfo {author}
  {\bibfnamefont {R.~J.}\ \bibnamefont {Poole}}, \bibinfo {author}
  {\bibfnamefont {T.}~\bibnamefont {Date}}, \bibinfo {author} {\bibfnamefont
  {T.}~\bibnamefont {Oshima}}, \bibinfo {author} {\bibfnamefont
  {J.}~\bibnamefont {Konishi}}, \bibinfo {author} {\bibfnamefont
  {K.}~\bibnamefont {Denda}}, \ and\ \bibinfo {author} {\bibfnamefont
  {M.}~\bibnamefont {Yoshida}},\ }\bibfield  {title} {\enquote {\bibinfo
  {title} {Evolution of the vacuolar {H}$^+$-atpase: {Implications} for the
  origin of {Eukaryotes}},}\ }\href@noop {} {\bibfield  {journal} {\bibinfo
  {journal} {Proc. Natl. Acad. Sci. USA}\ }\textbf {\bibinfo {volume} {86}},\
  \bibinfo {pages} {6661--6665} (\bibinfo {year} {1989})}\BibitemShut {NoStop}%
\bibitem [{\citenamefont {Iwabe}\ \emph {et~al.}(1989)\citenamefont {Iwabe},
  \citenamefont {Kuma}, \citenamefont {Hasegawa}, \citenamefont {Osawa},\ and\
  \citenamefont {Miyata}}]{BioIwabe1989}%
  \BibitemOpen
  \bibfield  {author} {\bibinfo {author} {\bibfnamefont {Naoyuki}\ \bibnamefont
  {Iwabe}}, \bibinfo {author} {\bibfnamefont {Kei-ichi}\ \bibnamefont {Kuma}},
  \bibinfo {author} {\bibfnamefont {Masami}\ \bibnamefont {Hasegawa}}, \bibinfo
  {author} {\bibfnamefont {Syozo}\ \bibnamefont {Osawa}}, \ and\ \bibinfo
  {author} {\bibfnamefont {Takashi}\ \bibnamefont {Miyata}},\ }\bibfield
  {title} {\enquote {\bibinfo {title} {Evolutionary relationship of
  archaebacteria, eubacteria, and eukaryotes inferred from phylogenetic trees
  of duplicated genes},}\ }\href@noop {} {\bibfield  {journal} {\bibinfo
  {journal} {Proceedings of the National Academy of Sciences}\ }\textbf
  {\bibinfo {volume} {86}},\ \bibinfo {pages} {9355--9359} (\bibinfo {year}
  {1989})}\BibitemShut {NoStop}%
\bibitem [{\citenamefont {Theobald}(2010)}]{BioTheobald2010}%
  \BibitemOpen
  \bibfield  {author} {\bibinfo {author} {\bibfnamefont {D.~L.}\ \bibnamefont
  {Theobald}},\ }\bibfield  {title} {\enquote {\bibinfo {title} {A formal test
  of the theory of universal common ancestry},}\ }\href@noop {} {\bibfield
  {journal} {\bibinfo  {journal} {Nature}\ }\textbf {\bibinfo {volume} {415}},\
  \bibinfo {pages} {219--223} (\bibinfo {year} {2010})}\BibitemShut {NoStop}%
\bibitem [{\citenamefont {Dagan}\ \emph {et~al.}(2010)\citenamefont {Dagan},
  \citenamefont {Roettger}, \citenamefont {Bryant},\ and\ \citenamefont
  {Martin}}]{dagan2010genome}%
  \BibitemOpen
  \bibfield  {author} {\bibinfo {author} {\bibfnamefont {Tal}\ \bibnamefont
  {Dagan}}, \bibinfo {author} {\bibfnamefont {Mayo}\ \bibnamefont {Roettger}},
  \bibinfo {author} {\bibfnamefont {David}\ \bibnamefont {Bryant}}, \ and\
  \bibinfo {author} {\bibfnamefont {William}\ \bibnamefont {Martin}},\
  }\bibfield  {title} {\enquote {\bibinfo {title} {Genome networks root the
  tree of life between prokaryotic domains},}\ }\href@noop {} {\bibfield
  {journal} {\bibinfo  {journal} {Genome Biology and Evolution}\ }\textbf
  {\bibinfo {volume} {2}},\ \bibinfo {pages} {379--392} (\bibinfo {year}
  {2010})}\BibitemShut {NoStop}%
\bibitem [{\citenamefont {Williams}\ \emph {et~al.}(2013)\citenamefont
  {Williams}, \citenamefont {Foster}, \citenamefont {Cox},\ and\ \citenamefont
  {Embley}}]{williams2013archaeal}%
  \BibitemOpen
  \bibfield  {author} {\bibinfo {author} {\bibfnamefont {Tom~A}\ \bibnamefont
  {Williams}}, \bibinfo {author} {\bibfnamefont {Peter~G}\ \bibnamefont
  {Foster}}, \bibinfo {author} {\bibfnamefont {Cymon~J}\ \bibnamefont {Cox}}, \
  and\ \bibinfo {author} {\bibfnamefont {T~Martin}\ \bibnamefont {Embley}},\
  }\bibfield  {title} {\enquote {\bibinfo {title} {An archaeal origin of
  eukaryotes supports only two primary domains of life},}\ }\href@noop {}
  {\bibfield  {journal} {\bibinfo  {journal} {Nature}\ }\textbf {\bibinfo
  {volume} {504}},\ \bibinfo {pages} {231--236} (\bibinfo {year}
  {2013})}\BibitemShut {NoStop}%
\bibitem [{\citenamefont {Glansdorff}\ \emph {et~al.}(2008)\citenamefont
  {Glansdorff}, \citenamefont {Xu}, \citenamefont {Labedan} \emph
  {et~al.}}]{glansdorff2008last}%
  \BibitemOpen
  \bibfield  {author} {\bibinfo {author} {\bibfnamefont {Nicolas}\ \bibnamefont
  {Glansdorff}}, \bibinfo {author} {\bibfnamefont {Ying}\ \bibnamefont {Xu}},
  \bibinfo {author} {\bibfnamefont {Bernard}\ \bibnamefont {Labedan}},  \emph
  {et~al.},\ }\bibfield  {title} {\enquote {\bibinfo {title} {The last
  universal common ancestor: emergence, constitution and genetic legacy of an
  elusive forerunner},}\ }\href@noop {} {\bibfield  {journal} {\bibinfo
  {journal} {Biol Direct}\ }\textbf {\bibinfo {volume} {3}},\ \bibinfo {pages}
  {56--125} (\bibinfo {year} {2008})}\BibitemShut {NoStop}%
\bibitem [{\citenamefont {Woese}\ and\ \citenamefont
  {Fox}(1977{\natexlab{b}})}]{WoeseFox_progenote1977}%
  \BibitemOpen
  \bibfield  {author} {\bibinfo {author} {\bibfnamefont {C.~R.}\ \bibnamefont
  {Woese}}\ and\ \bibinfo {author} {\bibfnamefont {G.~E.}\ \bibnamefont
  {Fox}},\ }\bibfield  {title} {\enquote {\bibinfo {title} {The concept of
  cellular evolution},}\ }\href@noop {} {\bibfield  {journal} {\bibinfo
  {journal} {J. Mol. Evol.}\ }\textbf {\bibinfo {volume} {10}},\ \bibinfo
  {pages} {1--6} (\bibinfo {year} {1977}{\natexlab{b}})}\BibitemShut {NoStop}%
\bibitem [{\citenamefont {Woese}(1998)}]{HGTWoese1998}%
  \BibitemOpen
  \bibfield  {author} {\bibinfo {author} {\bibfnamefont {C.~R.}\ \bibnamefont
  {Woese}},\ }\bibfield  {title} {\enquote {\bibinfo {title} {The universal
  ancestor},}\ }\href@noop {} {\bibfield  {journal} {\bibinfo  {journal} {Proc.
  Natl. Acad. Sci. USA}\ }\textbf {\bibinfo {volume} {95}},\ \bibinfo {pages}
  {6854--6859} (\bibinfo {year} {1998})}\BibitemShut {NoStop}%
\bibitem [{\citenamefont {Woese}(2002)}]{HGTWoese2002}%
  \BibitemOpen
  \bibfield  {author} {\bibinfo {author} {\bibfnamefont {C.~R.}\ \bibnamefont
  {Woese}},\ }\bibfield  {title} {\enquote {\bibinfo {title} {On the evolution
  of cells},}\ }\href@noop {} {\bibfield  {journal} {\bibinfo  {journal} {Proc.
  Natl. Acad. Sci. USA}\ }\textbf {\bibinfo {volume} {99}},\ \bibinfo {pages}
  {8742--8747} (\bibinfo {year} {2002})}\BibitemShut {NoStop}%
\bibitem [{\citenamefont {Woese}(2005)}]{HGTWoese2005}%
  \BibitemOpen
  \bibfield  {author} {\bibinfo {author} {\bibfnamefont {C.~R.}\ \bibnamefont
  {Woese}},\ }\bibfield  {title} {\enquote {\bibinfo {title} {Evolving
  biological organization},}\ }in\ \href@noop {} {\emph {\bibinfo {booktitle}
  {Microbial Phylogeny and Evolution: Concepts and Controversies}}},\ \bibinfo
  {editor} {edited by\ \bibinfo {editor} {\bibfnamefont {J.}~\bibnamefont
  {Sapp}}}\ (\bibinfo  {publisher} {Oxford University Press},\ \bibinfo
  {address} {Oxford, UK},\ \bibinfo {year} {2005})\ pp.\ \bibinfo {pages}
  {99--117}\BibitemShut {NoStop}%
\bibitem [{\citenamefont {Vetsigian}\ \emph {et~al.}(2006)\citenamefont
  {Vetsigian}, \citenamefont {C.Woese},\ and\ \citenamefont
  {Goldenfeld}}]{HGTVetsigian2006}%
  \BibitemOpen
  \bibfield  {author} {\bibinfo {author} {\bibfnamefont {K.}~\bibnamefont
  {Vetsigian}}, \bibinfo {author} {\bibnamefont {C.Woese}}, \ and\ \bibinfo
  {author} {\bibfnamefont {N.}~\bibnamefont {Goldenfeld}},\ }\bibfield  {title}
  {\enquote {\bibinfo {title} {Collective evolution and the genetic code},}\
  }\href@noop {} {\bibfield  {journal} {\bibinfo  {journal} {Proc. Natl. Acad.
  Sci. USA}\ }\textbf {\bibinfo {volume} {103}},\ \bibinfo {pages}
  {10696--10701} (\bibinfo {year} {2006})}\BibitemShut {NoStop}%
\bibitem [{\citenamefont {Haig}\ and\ \citenamefont
  {Hurst}(1991)}]{haig1991quantitative}%
  \BibitemOpen
  \bibfield  {author} {\bibinfo {author} {\bibfnamefont {David}\ \bibnamefont
  {Haig}}\ and\ \bibinfo {author} {\bibfnamefont {Laurence~D}\ \bibnamefont
  {Hurst}},\ }\bibfield  {title} {\enquote {\bibinfo {title} {A quantitative
  measure of error minimization in the genetic code},}\ }\href@noop {}
  {\bibfield  {journal} {\bibinfo  {journal} {Journal of Molecular Evolution}\
  }\textbf {\bibinfo {volume} {33}},\ \bibinfo {pages} {412--417} (\bibinfo
  {year} {1991})}\BibitemShut {NoStop}%
\bibitem [{\citenamefont {Freeland}\ and\ \citenamefont
  {Hurst}(1998)}]{freeland1998genetic}%
  \BibitemOpen
  \bibfield  {author} {\bibinfo {author} {\bibfnamefont {Stephen~J}\
  \bibnamefont {Freeland}}\ and\ \bibinfo {author} {\bibfnamefont {Laurence~D}\
  \bibnamefont {Hurst}},\ }\bibfield  {title} {\enquote {\bibinfo {title} {The
  genetic code is one in a million},}\ }\href@noop {} {\bibfield  {journal}
  {\bibinfo  {journal} {Journal of Molecular Evolution}\ }\textbf {\bibinfo
  {volume} {47}},\ \bibinfo {pages} {238--248} (\bibinfo {year}
  {1998})}\BibitemShut {NoStop}%
\bibitem [{\citenamefont {Butler}\ \emph {et~al.}(2009)\citenamefont {Butler},
  \citenamefont {Goldenfeld}, \citenamefont {Mathew},\ and\ \citenamefont
  {Luthey-Schulten}}]{butler2009extreme}%
  \BibitemOpen
  \bibfield  {author} {\bibinfo {author} {\bibfnamefont {Thomas}\ \bibnamefont
  {Butler}}, \bibinfo {author} {\bibfnamefont {Nigel}\ \bibnamefont
  {Goldenfeld}}, \bibinfo {author} {\bibfnamefont {Damien}\ \bibnamefont
  {Mathew}}, \ and\ \bibinfo {author} {\bibfnamefont {Zaida}\ \bibnamefont
  {Luthey-Schulten}},\ }\bibfield  {title} {\enquote {\bibinfo {title} {Extreme
  genetic code optimality from a molecular dynamics calculation of amino acid
  polar requirement},}\ }\href@noop {} {\bibfield  {journal} {\bibinfo
  {journal} {Physical Review E}\ }\textbf {\bibinfo {volume} {79}},\ \bibinfo
  {pages} {060901} (\bibinfo {year} {2009})}\BibitemShut {NoStop}%
\bibitem [{\citenamefont {Drossel}(2001)}]{ClassicDrossel2001}%
  \BibitemOpen
  \bibfield  {author} {\bibinfo {author} {\bibfnamefont {B.}~\bibnamefont
  {Drossel}},\ }\bibfield  {title} {\enquote {\bibinfo {title} {Biological
  evolution and statistical physics},}\ }\href@noop {} {\bibfield  {journal}
  {\bibinfo  {journal} {Adv. Phys.}\ }\textbf {\bibinfo {volume} {50}},\
  \bibinfo {pages} {209--295} (\bibinfo {year} {2001})}\BibitemShut {NoStop}%
\bibitem [{\citenamefont {Syvanen}(1985)}]{HGTSyvanen1985}%
  \BibitemOpen
  \bibfield  {author} {\bibinfo {author} {\bibfnamefont {M.}~\bibnamefont
  {Syvanen}},\ }\bibfield  {title} {\enquote {\bibinfo {title} {Cross-species
  gene transfer; implications for a new theory of evolution},}\ }\href@noop {}
  {\bibfield  {journal} {\bibinfo  {journal} {J. Theor. Biol.}\ }\textbf
  {\bibinfo {volume} {112}},\ \bibinfo {pages} {333--343} (\bibinfo {year}
  {1985})}\BibitemShut {NoStop}%
\bibitem [{\citenamefont {Syvanen}(1994)}]{HGTSyvanen1994}%
  \BibitemOpen
  \bibfield  {author} {\bibinfo {author} {\bibfnamefont {M.}~\bibnamefont
  {Syvanen}},\ }\bibfield  {title} {\enquote {\bibinfo {title} {Horizontal gene
  transfer: {Evidence} and possible consequences},}\ }\href@noop {} {\bibfield
  {journal} {\bibinfo  {journal} {Annu. Rev. Genet.}\ }\textbf {\bibinfo
  {volume} {28}},\ \bibinfo {pages} {237--261} (\bibinfo {year}
  {1994})}\BibitemShut {NoStop}%
\bibitem [{\citenamefont {Delsuc}\ \emph {et~al.}(2005)\citenamefont {Delsuc},
  \citenamefont {Brinkmann},\ and\ \citenamefont {Philippe}}]{BioDelsuc2005}%
  \BibitemOpen
  \bibfield  {author} {\bibinfo {author} {\bibfnamefont {F.}~\bibnamefont
  {Delsuc}}, \bibinfo {author} {\bibfnamefont {H.}~\bibnamefont {Brinkmann}}, \
  and\ \bibinfo {author} {\bibfnamefont {H.}~\bibnamefont {Philippe}},\
  }\bibfield  {title} {\enquote {\bibinfo {title} {Phylogenomics and the
  reconstruction of the tree of life},}\ }\href@noop {} {\bibfield  {journal}
  {\bibinfo  {journal} {Nat. Rev. Genet.}\ }\textbf {\bibinfo {volume} {6}},\
  \bibinfo {pages} {361--375} (\bibinfo {year} {2005})}\BibitemShut {NoStop}%
\bibitem [{\citenamefont {Thomas}\ and\ \citenamefont
  {Nielsen}(2005)}]{HGTThomas2005}%
  \BibitemOpen
  \bibfield  {author} {\bibinfo {author} {\bibfnamefont {C.~M.}\ \bibnamefont
  {Thomas}}\ and\ \bibinfo {author} {\bibfnamefont {K.~M.}\ \bibnamefont
  {Nielsen}},\ }\bibfield  {title} {\enquote {\bibinfo {title} {Mechanisms of,
  and barriers to, horizontal gene transfer between bacteria},}\ }\href@noop {}
  {\bibfield  {journal} {\bibinfo  {journal} {Nat. Rev. Microbiol.}\ }\textbf
  {\bibinfo {volume} {3}},\ \bibinfo {pages} {711--721} (\bibinfo {year}
  {2005})}\BibitemShut {NoStop}%
\bibitem [{\citenamefont {Anderson}(1966)}]{anderson1966possible}%
  \BibitemOpen
  \bibfield  {author} {\bibinfo {author} {\bibfnamefont {ES}~\bibnamefont
  {Anderson}},\ }\bibfield  {title} {\enquote {\bibinfo {title} {Possible
  importance of transfer factors in bacterial evolution},}\ }\href@noop {}
  {\bibfield  {journal} {\bibinfo  {journal} {Nature}\ }\textbf {\bibinfo
  {volume} {209}},\ \bibinfo {pages} {637--638} (\bibinfo {year}
  {1966})}\BibitemShut {NoStop}%
\bibitem [{\citenamefont {Anderson}(1970)}]{anderson1970evolutionary}%
  \BibitemOpen
  \bibfield  {author} {\bibinfo {author} {\bibfnamefont {Norman~G}\
  \bibnamefont {Anderson}},\ }\bibfield  {title} {\enquote {\bibinfo {title}
  {Evolutionary significance of virus infection},}\ }\href@noop {} {\bibfield
  {journal} {\bibinfo  {journal} {Nature}\ }\textbf {\bibinfo {volume} {227}},\
  \bibinfo {pages} {1346--1347} (\bibinfo {year} {1970})}\BibitemShut {NoStop}%
\bibitem [{\citenamefont {Sonea}(1988)}]{sonea1988bacterial}%
  \BibitemOpen
  \bibfield  {author} {\bibinfo {author} {\bibfnamefont {Sorin}\ \bibnamefont
  {Sonea}},\ }\bibfield  {title} {\enquote {\bibinfo {title} {A bacterial way
  of life},}\ }\href@noop {} {\bibfield  {journal} {\bibinfo  {journal}
  {Nature}\ }\textbf {\bibinfo {volume} {331}},\ \bibinfo {pages} {216}
  (\bibinfo {year} {1988})}\BibitemShut {NoStop}%
\bibitem [{\citenamefont {Kurland}\ \emph {et~al.}(2003)\citenamefont
  {Kurland}, \citenamefont {Canback},\ and\ \citenamefont
  {Berg}}]{HGTKurland2003}%
  \BibitemOpen
  \bibfield  {author} {\bibinfo {author} {\bibfnamefont {C.~G.}\ \bibnamefont
  {Kurland}}, \bibinfo {author} {\bibfnamefont {B.}~\bibnamefont {Canback}}, \
  and\ \bibinfo {author} {\bibfnamefont {O.~G.}\ \bibnamefont {Berg}},\
  }\bibfield  {title} {\enquote {\bibinfo {title} {Horizontal gene transfer:
  {A} critical view},}\ }\href@noop {} {\bibfield  {journal} {\bibinfo
  {journal} {Proc. Natl. Acad. Sci. U. S. A.}\ }\textbf {\bibinfo {volume}
  {100}},\ \bibinfo {pages} {9658--9662} (\bibinfo {year} {2003})}\BibitemShut
  {NoStop}%
\bibitem [{\citenamefont {Dagan}\ and\ \citenamefont
  {Martin}(2006)}]{HGTDagan2006}%
  \BibitemOpen
  \bibfield  {author} {\bibinfo {author} {\bibfnamefont {T.}~\bibnamefont
  {Dagan}}\ and\ \bibinfo {author} {\bibfnamefont {W.}~\bibnamefont {Martin}},\
  }\bibfield  {title} {\enquote {\bibinfo {title} {The tree of one percent},}\
  }\href@noop {} {\bibfield  {journal} {\bibinfo  {journal} {Genome Biol.}\
  }\textbf {\bibinfo {volume} {7}},\ \bibinfo {pages} {118} (\bibinfo {year}
  {2006})}\BibitemShut {NoStop}%
\bibitem [{\citenamefont {Takeuchi}\ \emph {et~al.}(2014)\citenamefont
  {Takeuchi}, \citenamefont {Kaneko},\ and\ \citenamefont
  {Koonin}}]{takeuchi2014horizontal}%
  \BibitemOpen
  \bibfield  {author} {\bibinfo {author} {\bibfnamefont {Nobuto}\ \bibnamefont
  {Takeuchi}}, \bibinfo {author} {\bibfnamefont {Kunihiko}\ \bibnamefont
  {Kaneko}}, \ and\ \bibinfo {author} {\bibfnamefont {Eugene~V}\ \bibnamefont
  {Koonin}},\ }\bibfield  {title} {\enquote {\bibinfo {title} {Horizontal gene
  transfer can rescue prokaryotes from mullerÕs ratchet: Benefit of dna from
  dead cells and population subdivision},}\ }\href@noop {} {\bibfield
  {journal} {\bibinfo  {journal} {G3: Genes | Genomes | Genetics}\ }\textbf
  {\bibinfo {volume} {4}},\ \bibinfo {pages} {325--339} (\bibinfo {year}
  {2014})}\BibitemShut {NoStop}%
\bibitem [{\citenamefont {Cohan}(2002)}]{cohan2002bacterial}%
  \BibitemOpen
  \bibfield  {author} {\bibinfo {author} {\bibfnamefont {Frederick~M}\
  \bibnamefont {Cohan}},\ }\bibfield  {title} {\enquote {\bibinfo {title} {What
  are bacterial species?}}\ }\href@noop {} {\bibfield  {journal} {\bibinfo
  {journal} {Annual Reviews in Microbiology}\ }\textbf {\bibinfo {volume}
  {56}},\ \bibinfo {pages} {457--487} (\bibinfo {year} {2002})}\BibitemShut
  {NoStop}%
\bibitem [{\citenamefont {Lawrence}(2002)}]{lawrence2002gene}%
  \BibitemOpen
  \bibfield  {author} {\bibinfo {author} {\bibfnamefont {Jeffrey~G}\
  \bibnamefont {Lawrence}},\ }\bibfield  {title} {\enquote {\bibinfo {title}
  {Gene transfer in bacteria: speciation without species?}}\ }\href@noop {}
  {\bibfield  {journal} {\bibinfo  {journal} {Theoretical Population Biology}\
  }\textbf {\bibinfo {volume} {61}},\ \bibinfo {pages} {449--460} (\bibinfo
  {year} {2002})}\BibitemShut {NoStop}%
\bibitem [{\citenamefont {Gevers}\ \emph {et~al.}(2005)\citenamefont {Gevers},
  \citenamefont {Cohan}, \citenamefont {Lawrence}, \citenamefont {Spratt},
  \citenamefont {Coenye}, \citenamefont {Feil}, \citenamefont {Stackebrandt},
  \citenamefont {Van~de Peer}, \citenamefont {Vandamme}, \citenamefont
  {Thompson} \emph {et~al.}}]{gevers2005re}%
  \BibitemOpen
  \bibfield  {author} {\bibinfo {author} {\bibfnamefont {Dirk}\ \bibnamefont
  {Gevers}}, \bibinfo {author} {\bibfnamefont {Frederick~M}\ \bibnamefont
  {Cohan}}, \bibinfo {author} {\bibfnamefont {Jeffrey~G}\ \bibnamefont
  {Lawrence}}, \bibinfo {author} {\bibfnamefont {Brian~G}\ \bibnamefont
  {Spratt}}, \bibinfo {author} {\bibfnamefont {Tom}\ \bibnamefont {Coenye}},
  \bibinfo {author} {\bibfnamefont {Edward~J}\ \bibnamefont {Feil}}, \bibinfo
  {author} {\bibfnamefont {Erko}\ \bibnamefont {Stackebrandt}}, \bibinfo
  {author} {\bibfnamefont {Yves}\ \bibnamefont {Van~de Peer}}, \bibinfo
  {author} {\bibfnamefont {Peter}\ \bibnamefont {Vandamme}}, \bibinfo {author}
  {\bibfnamefont {Fabiano~L}\ \bibnamefont {Thompson}},  \emph {et~al.},\
  }\bibfield  {title} {\enquote {\bibinfo {title} {Re-evaluating prokaryotic
  species},}\ }\href@noop {} {\bibfield  {journal} {\bibinfo  {journal} {Nature
  Reviews Microbiology}\ }\textbf {\bibinfo {volume} {3}},\ \bibinfo {pages}
  {733--739} (\bibinfo {year} {2005})}\BibitemShut {NoStop}%
\bibitem [{\citenamefont {Konstantinidis}\ and\ \citenamefont
  {Tiedje}(2005)}]{konstantinidis2005genomic}%
  \BibitemOpen
  \bibfield  {author} {\bibinfo {author} {\bibfnamefont {Konstantinos~T}\
  \bibnamefont {Konstantinidis}}\ and\ \bibinfo {author} {\bibfnamefont
  {James~M}\ \bibnamefont {Tiedje}},\ }\bibfield  {title} {\enquote {\bibinfo
  {title} {Genomic insights that advance the species definition for
  prokaryotes},}\ }\href@noop {} {\bibfield  {journal} {\bibinfo  {journal}
  {Proceedings of the National Academy of Sciences USA}\ }\textbf {\bibinfo
  {volume} {102}},\ \bibinfo {pages} {2567--2572} (\bibinfo {year}
  {2005})}\BibitemShut {NoStop}%
\bibitem [{\citenamefont {Staley}(2006)}]{staley2006bacterial}%
  \BibitemOpen
  \bibfield  {author} {\bibinfo {author} {\bibfnamefont {James~T}\ \bibnamefont
  {Staley}},\ }\bibfield  {title} {\enquote {\bibinfo {title} {The bacterial
  species dilemma and the genomic--phylogenetic species concept},}\ }\href@noop
  {} {\bibfield  {journal} {\bibinfo  {journal} {Philosophical Transactions of
  the Royal Society B: Biological Sciences}\ }\textbf {\bibinfo {volume}
  {361}},\ \bibinfo {pages} {1899--1909} (\bibinfo {year} {2006})}\BibitemShut
  {NoStop}%
\bibitem [{\citenamefont {Zhaxybayeva}\ and\ \citenamefont
  {Doolittle}(2011)}]{HGTDoolittle2011}%
  \BibitemOpen
  \bibfield  {author} {\bibinfo {author} {\bibfnamefont {O.}~\bibnamefont
  {Zhaxybayeva}}\ and\ \bibinfo {author} {\bibfnamefont {W.~F.}\ \bibnamefont
  {Doolittle}},\ }\bibfield  {title} {\enquote {\bibinfo {title} {Lateral gene
  transfer},}\ }\href@noop {} {\bibfield  {journal} {\bibinfo  {journal} {Curr.
  Biol.}\ }\textbf {\bibinfo {volume} {21}},\ \bibinfo {pages} {R242--R246}
  (\bibinfo {year} {2011})}\BibitemShut {NoStop}%
\bibitem [{\citenamefont {Goldenfeld}\ and\ \citenamefont
  {Woese}(2007)}]{HGTGoldenfeld2007}%
  \BibitemOpen
  \bibfield  {author} {\bibinfo {author} {\bibfnamefont {N.}~\bibnamefont
  {Goldenfeld}}\ and\ \bibinfo {author} {\bibfnamefont {C.}~\bibnamefont
  {Woese}},\ }\bibfield  {title} {\enquote {\bibinfo {title} {Biology's next
  revolution},}\ }\href@noop {} {\bibfield  {journal} {\bibinfo  {journal}
  {Nature}\ }\textbf {\bibinfo {volume} {445}},\ \bibinfo {pages} {369}
  (\bibinfo {year} {2007})}\BibitemShut {NoStop}%
\bibitem [{\citenamefont {Dyson}(2007)}]{HGTDysonNewYorkReview2007}%
  \BibitemOpen
  \bibfield  {author} {\bibinfo {author} {\bibfnamefont {F.}~\bibnamefont
  {Dyson}},\ }\bibfield  {title} {\enquote {\bibinfo {title} {Our biotech
  future},}\ }\href@noop {} {\bibfield  {journal} {\bibinfo  {journal} {New
  York Review of Books}\ }\textbf {\bibinfo {volume} {July 19}} (\bibinfo
  {year} {2007})}\BibitemShut {NoStop}%
\bibitem [{\citenamefont {Taylor}\ \emph {et~al.}(2004)\citenamefont {Taylor},
  \citenamefont {Fudenberg}, \citenamefont {Sasaki},\ and\ \citenamefont
  {Nowak}}]{ClassicTaylor2004}%
  \BibitemOpen
  \bibfield  {author} {\bibinfo {author} {\bibfnamefont {C.}~\bibnamefont
  {Taylor}}, \bibinfo {author} {\bibfnamefont {D.}~\bibnamefont {Fudenberg}},
  \bibinfo {author} {\bibfnamefont {A.}~\bibnamefont {Sasaki}}, \ and\ \bibinfo
  {author} {\bibfnamefont {M.~A.}\ \bibnamefont {Nowak}},\ }\bibfield  {title}
  {\enquote {\bibinfo {title} {Evolutionary game theory in finite
  populations},}\ }\href@noop {} {\bibfield  {journal} {\bibinfo  {journal}
  {Bull. Math. Biol.}\ }\textbf {\bibinfo {volume} {66}},\ \bibinfo {pages}
  {1621--1644} (\bibinfo {year} {2004})}\BibitemShut {NoStop}%
\bibitem [{\citenamefont {Traulsen}\ \emph {et~al.}(2005)\citenamefont
  {Traulsen}, \citenamefont {Claussen},\ and\ \citenamefont
  {Hauert}}]{ClassicTraulsen2005}%
  \BibitemOpen
  \bibfield  {author} {\bibinfo {author} {\bibfnamefont {A.}~\bibnamefont
  {Traulsen}}, \bibinfo {author} {\bibfnamefont {J.~C.}\ \bibnamefont
  {Claussen}}, \ and\ \bibinfo {author} {\bibfnamefont {C.}~\bibnamefont
  {Hauert}},\ }\bibfield  {title} {\enquote {\bibinfo {title} {Coevolutionary
  dynamics: {From} finite to infinite populations},}\ }\href@noop {} {\bibfield
   {journal} {\bibinfo  {journal} {Phys. Rev. Lett.}\ }\textbf {\bibinfo
  {volume} {95}},\ \bibinfo {pages} {238701} (\bibinfo {year}
  {2005})}\BibitemShut {NoStop}%
\bibitem [{\citenamefont {Nowak}(2006)}]{BasicsNowak2006}%
  \BibitemOpen
  \bibfield  {author} {\bibinfo {author} {\bibfnamefont {M.~A.}\ \bibnamefont
  {Nowak}},\ }\href@noop {} {\emph {\bibinfo {title} {Evolutionary Dynamics}}}\
  (\bibinfo  {publisher} {Harvard University Press},\ \bibinfo {address}
  {London, UK},\ \bibinfo {year} {2006})\BibitemShut {NoStop}%
\bibitem [{\citenamefont {Durrett}\ and\ \citenamefont
  {Schmidt}(2008)}]{BioDurrett2008}%
  \BibitemOpen
  \bibfield  {author} {\bibinfo {author} {\bibfnamefont {R.}~\bibnamefont
  {Durrett}}\ and\ \bibinfo {author} {\bibfnamefont {D.}~\bibnamefont
  {Schmidt}},\ }\bibfield  {title} {\enquote {\bibinfo {title} {Waiting for two
  mutations: {With} applications to regulatory sequence evolution and the
  limits of {Darwinian} evolution},}\ }\href@noop {} {\bibfield  {journal}
  {\bibinfo  {journal} {Genetics}\ }\textbf {\bibinfo {volume} {180}},\
  \bibinfo {pages} {1501--1509} (\bibinfo {year} {2008})}\BibitemShut {NoStop}%
\bibitem [{\citenamefont {Arnoldt}\ \emph {et~al.}(2012)\citenamefont
  {Arnoldt}, \citenamefont {Timme},\ and\ \citenamefont
  {Grosskinsky}}]{ClassicArnoldt2012}%
  \BibitemOpen
  \bibfield  {author} {\bibinfo {author} {\bibfnamefont {H.}~\bibnamefont
  {Arnoldt}}, \bibinfo {author} {\bibfnamefont {M.}~\bibnamefont {Timme}}, \
  and\ \bibinfo {author} {\bibfnamefont {S.}~\bibnamefont {Grosskinsky}},\
  }\bibfield  {title} {\enquote {\bibinfo {title} {Frequency dependent fitness
  induces multistability in coevolutionary dynamics},}\ }\href@noop {}
  {\bibfield  {journal} {\bibinfo  {journal} {J. R. Soc. Interface}\ }\textbf
  {\bibinfo {volume} {9}},\ \bibinfo {pages} {3387--3396} (\bibinfo {year}
  {2012})}\BibitemShut {NoStop}%
\bibitem [{\citenamefont {Leisner}\ \emph {et~al.}(2008)\citenamefont
  {Leisner}, \citenamefont {Stingl}, \citenamefont {Frey},\ and\ \citenamefont
  {Maier}}]{HGTLeisner2008}%
  \BibitemOpen
  \bibfield  {author} {\bibinfo {author} {\bibfnamefont {M.}~\bibnamefont
  {Leisner}}, \bibinfo {author} {\bibfnamefont {K.}~\bibnamefont {Stingl}},
  \bibinfo {author} {\bibfnamefont {E.}~\bibnamefont {Frey}}, \ and\ \bibinfo
  {author} {\bibfnamefont {B.}~\bibnamefont {Maier}},\ }\bibfield  {title}
  {\enquote {\bibinfo {title} {Stochastic switching to competence},}\
  }\href@noop {} {\bibfield  {journal} {\bibinfo  {journal} {Curr. Opin.
  Microbiol.}\ }\textbf {\bibinfo {volume} {11}},\ \bibinfo {pages} {553--559}
  (\bibinfo {year} {2008})}\BibitemShut {NoStop}%
\bibitem [{\citenamefont {Moran}\ and\ \citenamefont
  {Jarvik}(2010)}]{MoranScience2010}%
  \BibitemOpen
  \bibfield  {author} {\bibinfo {author} {\bibfnamefont {N.~A.}\ \bibnamefont
  {Moran}}\ and\ \bibinfo {author} {\bibfnamefont {T.}~\bibnamefont {Jarvik}},\
  }\bibfield  {title} {\enquote {\bibinfo {title} {Lateral transfer of genes
  from fungi underlies carotenoid production in aphids},}\ }\href@noop {}
  {\bibfield  {journal} {\bibinfo  {journal} {Science}\ }\textbf {\bibinfo
  {volume} {328}},\ \bibinfo {pages} {624--627} (\bibinfo {year}
  {2010})}\BibitemShut {NoStop}%
\bibitem [{\citenamefont {Mayer}\ \emph {et~al.}(2011)\citenamefont {Mayer},
  \citenamefont {Schuster}, \citenamefont {Bartelmes}, \citenamefont
  {Dieterich},\ and\ \citenamefont {Sommer}}]{MayerBMCEvolBiol2011}%
  \BibitemOpen
  \bibfield  {author} {\bibinfo {author} {\bibfnamefont {W.~E.}\ \bibnamefont
  {Mayer}}, \bibinfo {author} {\bibfnamefont {L.~N.}\ \bibnamefont {Schuster}},
  \bibinfo {author} {\bibfnamefont {G.}~\bibnamefont {Bartelmes}}, \bibinfo
  {author} {\bibfnamefont {C.}~\bibnamefont {Dieterich}}, \ and\ \bibinfo
  {author} {\bibfnamefont {R.~J.}\ \bibnamefont {Sommer}},\ }\bibfield  {title}
  {\enquote {\bibinfo {title} {Horizontal gene transfer of microbial cellulases
  into nematode genomes is associated with functional assimilation and gene
  turnover},}\ }\href@noop {} {\bibfield  {journal} {\bibinfo  {journal} {BMC
  Evol. Biol.}\ }\textbf {\bibinfo {volume} {11}},\ \bibinfo {pages} {13}
  (\bibinfo {year} {2011})}\BibitemShut {NoStop}%
\bibitem [{\citenamefont {Vogan}\ and\ \citenamefont
  {Higgs}(2011)}]{HGTVogan2011}%
  \BibitemOpen
  \bibfield  {author} {\bibinfo {author} {\bibfnamefont {A.~A.}\ \bibnamefont
  {Vogan}}\ and\ \bibinfo {author} {\bibfnamefont {P.~G.}\ \bibnamefont
  {Higgs}},\ }\bibfield  {title} {\enquote {\bibinfo {title} {The advantages
  and disadvantages of horizontal gene transfer and the emergence of the first
  species},}\ }\href@noop {} {\bibfield  {journal} {\bibinfo  {journal} {Biol.
  Direct}\ }\textbf {\bibinfo {volume} {6}},\ \bibinfo {pages} {1} (\bibinfo
  {year} {2011})}\BibitemShut {NoStop}%
\bibitem [{\citenamefont {Fuentes}\ \emph {et~al.}(2014)\citenamefont
  {Fuentes}, \citenamefont {Stegemann}, \citenamefont {Golczyk}, \citenamefont
  {Karcher},\ and\ \citenamefont {Bock}}]{HGTFuentes2014}%
  \BibitemOpen
  \bibfield  {author} {\bibinfo {author} {\bibfnamefont {I.}~\bibnamefont
  {Fuentes}}, \bibinfo {author} {\bibfnamefont {S.}~\bibnamefont {Stegemann}},
  \bibinfo {author} {\bibfnamefont {H.}~\bibnamefont {Golczyk}}, \bibinfo
  {author} {\bibfnamefont {D.}~\bibnamefont {Karcher}}, \ and\ \bibinfo
  {author} {\bibfnamefont {R.}~\bibnamefont {Bock}},\ }\bibfield  {title}
  {\enquote {\bibinfo {title} {Horizontal genome transfer as an asexual path to
  the formation of new species},}\ }\href@noop {} {\bibfield  {journal}
  {\bibinfo  {journal} {Nature}\ } (\bibinfo {year} {2014})}\BibitemShut
  {NoStop}%
\bibitem [{\citenamefont {Wylie}\ \emph {et~al.}(2010)\citenamefont {Wylie},
  \citenamefont {Trout}, \citenamefont {Kessler},\ and\ \citenamefont
  {Levine}}]{HGTWylie2010}%
  \BibitemOpen
  \bibfield  {author} {\bibinfo {author} {\bibfnamefont {C.~S.}\ \bibnamefont
  {Wylie}}, \bibinfo {author} {\bibfnamefont {A.~D.}\ \bibnamefont {Trout}},
  \bibinfo {author} {\bibfnamefont {D.~A.}\ \bibnamefont {Kessler}}, \ and\
  \bibinfo {author} {\bibfnamefont {H.}~\bibnamefont {Levine}},\ }\bibfield
  {title} {\enquote {\bibinfo {title} {Optimal strategy for competence
  differentiation in bacteria},}\ }\href@noop {} {\bibfield  {journal}
  {\bibinfo  {journal} {PLoS Genetics}\ }\textbf {\bibinfo {volume} {6}}
  (\bibinfo {year} {2010})},\ \bibinfo {note} {doi:
  10.1371/journal.pgen.1001108}\BibitemShut {NoStop}%
\bibitem [{\citenamefont {Wilke}\ \emph {et~al.}(2001)\citenamefont {Wilke},
  \citenamefont {Ronnewinkel},\ and\ \citenamefont
  {Martinez}}]{ClassicWilke2001}%
  \BibitemOpen
  \bibfield  {author} {\bibinfo {author} {\bibfnamefont {C.~O.}\ \bibnamefont
  {Wilke}}, \bibinfo {author} {\bibfnamefont {C.}~\bibnamefont {Ronnewinkel}},
  \ and\ \bibinfo {author} {\bibfnamefont {T.}~\bibnamefont {Martinez}},\
  }\bibfield  {title} {\enquote {\bibinfo {title} {Dynamic fitness landscapes
  in molecular evolution},}\ }\href@noop {} {\bibfield  {journal} {\bibinfo
  {journal} {Phys. Rep.}\ }\textbf {\bibinfo {volume} {349}},\ \bibinfo {pages}
  {395--446} (\bibinfo {year} {2001})}\BibitemShut {NoStop}%
\bibitem [{\citenamefont {Moran}(1962)}]{ModelMoran1962}%
  \BibitemOpen
  \bibfield  {author} {\bibinfo {author} {\bibfnamefont {P.~A.~P.}\
  \bibnamefont {Moran}},\ }\href@noop {} {\emph {\bibinfo {title} {The
  Statistical Processes of Evolutionary Theory}}}\ (\bibinfo  {publisher}
  {Clarendon Press},\ \bibinfo {address} {Oxford, UK},\ \bibinfo {year}
  {1962})\BibitemShut {NoStop}%
\bibitem [{\citenamefont {Leisner}\ \emph {et~al.}(2009)\citenamefont
  {Leisner}, \citenamefont {Kuhr}, \citenamefont {R\"{a}dler}, \citenamefont
  {Frey},\ and\ \citenamefont {Maier}}]{HGTLeisner2009}%
  \BibitemOpen
  \bibfield  {author} {\bibinfo {author} {\bibfnamefont {M.}~\bibnamefont
  {Leisner}}, \bibinfo {author} {\bibfnamefont {J.-T.}\ \bibnamefont {Kuhr}},
  \bibinfo {author} {\bibfnamefont {J.~O.}\ \bibnamefont {R\"{a}dler}},
  \bibinfo {author} {\bibfnamefont {E.}~\bibnamefont {Frey}}, \ and\ \bibinfo
  {author} {\bibfnamefont {B.}~\bibnamefont {Maier}},\ }\bibfield  {title}
  {\enquote {\bibinfo {title} {Kinetics of genetic switching into the state of
  bacterial competence},}\ }\href@noop {} {\bibfield  {journal} {\bibinfo
  {journal} {Biophys. J.}\ }\textbf {\bibinfo {volume} {96}},\ \bibinfo {pages}
  {1178--1188} (\bibinfo {year} {2009})}\BibitemShut {NoStop}%
\bibitem [{\citenamefont {Kovacs}\ \emph {et~al.}(2009)\citenamefont {Kovacs},
  \citenamefont {Smits}, \citenamefont {Mironczuk},\ and\ \citenamefont
  {Kuipers}}]{HGTKovacs2009}%
  \BibitemOpen
  \bibfield  {author} {\bibinfo {author} {\bibfnamefont {A.~T.}\ \bibnamefont
  {Kovacs}}, \bibinfo {author} {\bibfnamefont {W.~K.}\ \bibnamefont {Smits}},
  \bibinfo {author} {\bibfnamefont {A.~M.}\ \bibnamefont {Mironczuk}}, \ and\
  \bibinfo {author} {\bibfnamefont {O.~P.}\ \bibnamefont {Kuipers}},\
  }\bibfield  {title} {\enquote {\bibinfo {title} {Ubiquitous late competence
  genes in bacillus species indicate the presence of functional dna uptake
  machineries},}\ }\href@noop {} {\bibfield  {journal} {\bibinfo  {journal}
  {Environ. Microbiol.}\ }\textbf {\bibinfo {volume} {11}},\ \bibinfo {pages}
  {1911--1922} (\bibinfo {year} {2009})}\BibitemShut {NoStop}%
\bibitem [{\citenamefont {H\"{a}nggi}\ \emph {et~al.}(1990)\citenamefont
  {H\"{a}nggi}, \citenamefont {Talkner},\ and\ \citenamefont
  {Borkovec}}]{ChemistryHaenggi1990}%
  \BibitemOpen
  \bibfield  {author} {\bibinfo {author} {\bibfnamefont {P.}~\bibnamefont
  {H\"{a}nggi}}, \bibinfo {author} {\bibfnamefont {P.}~\bibnamefont {Talkner}},
  \ and\ \bibinfo {author} {\bibfnamefont {M.}~\bibnamefont {Borkovec}},\
  }\bibfield  {title} {\enquote {\bibinfo {title} {Reaction rate theory:
  {Fifty} years after {Kramers}},}\ }\href@noop {} {\bibfield  {journal}
  {\bibinfo  {journal} {Rev. Mod. Phys.}\ }\textbf {\bibinfo {volume} {62}},\
  \bibinfo {pages} {251--342} (\bibinfo {year} {1990})}\BibitemShut {NoStop}%
\bibitem [{\citenamefont {Eigen}(1971)}]{ClassicEigen1971}%
  \BibitemOpen
  \bibfield  {author} {\bibinfo {author} {\bibfnamefont {M.}~\bibnamefont
  {Eigen}},\ }\bibfield  {title} {\enquote {\bibinfo {title} {Selforganization
  of matter and the evolution of biological macromolecules},}\ }\href@noop {}
  {\bibfield  {journal} {\bibinfo  {journal} {Die Naturwissenschaften}\
  }\textbf {\bibinfo {volume} {58}},\ \bibinfo {pages} {465--523} (\bibinfo
  {year} {1971})}\BibitemShut {NoStop}%
\bibitem [{\citenamefont {Eigen}\ and\ \citenamefont
  {Schuster}(1979)}]{ClassicEigen1979}%
  \BibitemOpen
  \bibfield  {author} {\bibinfo {author} {\bibfnamefont {M.}~\bibnamefont
  {Eigen}}\ and\ \bibinfo {author} {\bibfnamefont {P.}~\bibnamefont
  {Schuster}},\ }\href@noop {} {\emph {\bibinfo {title} {The Hypercycle: {A}
  Principle of Natural Self-Organization}}}\ (\bibinfo  {publisher}
  {Springer-Verlag},\ \bibinfo {address} {Berlin, Germany},\ \bibinfo {year}
  {1979})\BibitemShut {NoStop}%
\end{thebibliography}%

\end{document}